# Hydrogen Bonding in Ionic Liquids Probed by Linear and Nonlinear Vibrational Spectroscopy

C. Roth[1], S. Chatzipapadopoulos[2], D. Kerlé[1], F. Friedriszik[2], M. Lütgens[2], S. Lochbrunner[2], O. Kühn[3] and R. Ludwig[1,4]

[1] Institut für Chemie, Abteilung Physikalische Chemie, Universität Rostock, Dr.-Lorenz-Weg 1, D-18051 Rostock, Germany

[2] Institut für Physik, Universität Rostock, Universitätsplatz 3, D-18051 Rostock, Germany

[3] Institut für Physik, Universität Rostock, Wismarsche Str, 43-45, D-18057 Rostock, Germany

[4] Leibniz-Institut für Katalyse an der Universität Rostock, A.-Einstein-Str. 29a, 18059 Rostock, Germany

**Abstract**

Three imidazolium-based ionic liquids of the type $[C_n\text{mim}][NTf_2]$ with varying alkyl chain lengths (n = 1, 2 and 8) at the 1 position of the imidazolium ring were studied applying IR, linear Raman, and multiplex CARS spectroscopy. The focus has been on the CH-stretching region of the imidazolium ring, which is supposed to carry information about a possible hydrogen bonding network in the ionic liquid. The measurements are compared to calculations of the corresponding anharmonic vibrational spectra for a cluster of $[C_2\text{mim}][NTf_2]$ consisting of four ion pairs. The results support the hypothesis of weak hydrogen bonding involving the C(4)-H and C(5)-H groups and somewhat stronger hydrogen bonds of the C(2)-H groups.

## 1. Introduction

Ionic Liquids (ILs) have unique and fascinating properties providing a remarkable opportunity for new science and technology.[1-7] These liquid materials offer a wide range of possible applications as solvents for reaction and material processing, as extraction media or as working fluids in mechanical processes. The physical properties and the solvent-behavior of ILs are the key features for any application. The structure and properties of these Coulomb systems are mainly determined by the type and strength of the intermolecular interactions between anions and cations. In particular the subtle balance between Coulomb forces, hydrogen bonds (HBs), and dispersion forces is of great importance for understanding ILs. It is assumed that in particular hydrogen bonding plays an important role for the properties and reaction dynamics of these Coulomb systems.[8-11] Strong evidence for the existence of hydrogen bonding is provided by X-ray diffraction, mid-infrared, and NMR spectroscopy. Observed indications for hydrogen bonding in imidazolium-based ILs are shorter C-H⋯anion distances, red shifted C-H stretching frequencies and downfield shifted C-H proton chemical shifts.[12-22] However, it has been argued that the corresponding signatures in the infrared (IR) spectra can also result from other contributions and that hydrogen bonding is not essential for understanding the properties of ILs.[23-27] Strength and properties of the anion-cation interaction including hydrogen bonding have been also studied by theoretical methods.[28-31]

In particular, the C-H stretching vibrations in imidazolium-based ILs should give some information about the existence and strength of hydrogen bonding in this Coulomb system. So far, the interpretation of the C-H region is highly controversial.[32,33] Exemplarily, let us consider ILs of the type [1-alkyl-3-methyl imidazolium][bis(trifluoromethanesulfonyl)imide] (see Scheme 1) with the acronyms [$C_n$mim][$NTf_2$] where n = 1, 2, or 8 denotes the length of the alkyl chain at the 1 position of the imidazolium ring.

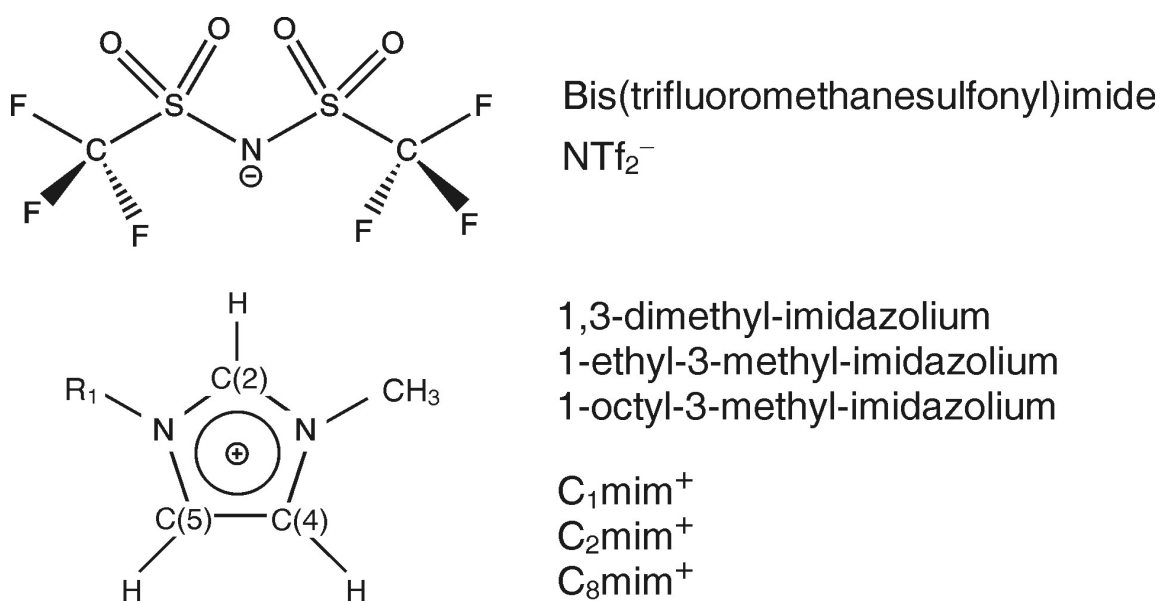

**Scheme 1:** Nomenclature used for [1-alkyl-3-methyl imidazolium][$NTf_2$] ionic liquids investigated in this work.

Their IR spectra of the C-H stretching region show contributions between 2800 cm$^{-1}$ and 3000 cm$^{-1}$, which can be clearly referred to the $CH_2$ and $CH_3$ stretching vibrations of the alkyl groups at the nitrogen atoms of the imidazolium ring. For the adjacent frequency range between 3100 cm$^{-1}$ and 3200 cm$^{-1}$ two main interpretations are given (Scheme 2). Grondin et al.[32] claim that the high frequency contributions at 3160±15 cm$^{-1}$ can be assigned exclusively to all C-H stretching vibrations of the imidazolium cation (C(2)-H, C(4)-H and C(5)-H), whereas the absorption at 3120±15 cm$^{-1}$ results, as indicated in Scheme 2, from overtones (2 R$_1$, 2 R$_2$) and combination tones (R$_1$ + R$_2$) of two in-plane ring vibrations R$_1$ and R$_2$ which form Fermi resonances with the C-H stretching vibrations. Albeit Fermi resonances of such combination and overtone transitions might contribute in the region at 3120cm$^{-1}$; Ludwig et al. claim that only a single vibrational band at 3160 cm$^{-1}$ cannot account for the three C(2)-H, C(4)-H and C(5)-H stretches.[33] They suggest that the band can be assigned to the C(4)-H and C(5)-H stretches, whereas the C(2)-H vibrational mode is shifted by about 50 cm$^{-1}$ to lower frequencies due to its stronger acidic character and thus falls into the frequency range of the overtones and combination tones. A redshift of the C(2)-H stretching frequency relative to those of C(4)-H and C(5)-H is in accordance with a stronger NMR downfield proton chemical shifts of about 1 ppm for C(2)-H compared to C(4,5)-H.[21] A further indication for this notion has been already observed by Grondin et al..[32] In isotopic substitution experiments they could record the C(2)-D, C(4)-D and C(5)-D stretching vibrational modes in the frequency range between 2250 cm$^{-1}$ and 2400 cm$^{-1}$. The benefit of these experiments is that this frequency range is not overcrowded by overtones and combination tones of the imidazolium ring. Grondin et al. observed the C(2)-D vibrational band at 2350-2355 cm$^{-1}$ and the C(4,5)-D vibrational bands at 2388-2393 cm$^{-1}$. Thus, the C(2)-D frequencies are clearly red-shifted by 38 cm$^{-1}$. Following the equation of the simple harmonic oscillator and taking the difference in the reduced masses into account (Grondin et al. estimated an isotopic ratio of 1.33), a frequency shift of $\Delta\nu = 50.5$ cm$^{-1}$ for the vibrational bands of the C-H bonds is expected in agreement with the suggestions by Ludwig et al..[27,33]

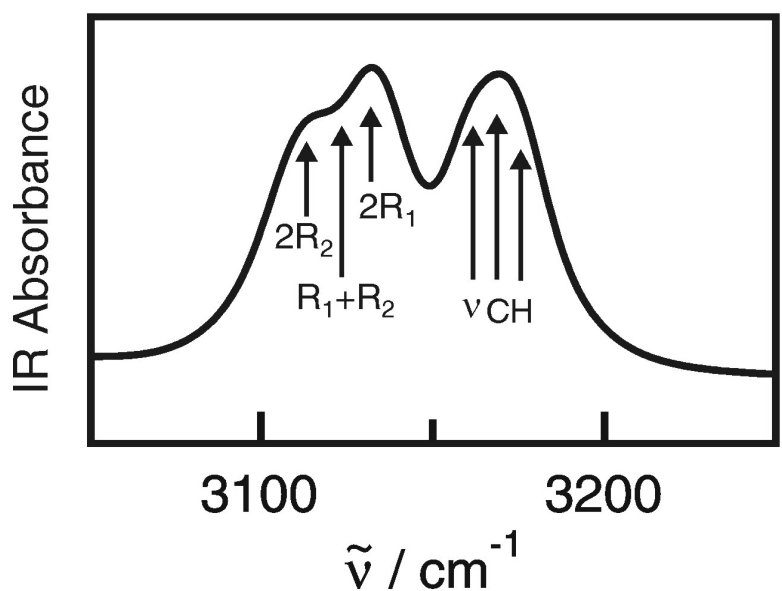

**Scheme 2:** Interpretation of the frequency region of C-H stretching vibrations in imidazolium-based ionic liquids as suggested by Grondin et al..[32]

To conclude, in the frequency range between 3100 cm$^{-1}$ and 3200 cm$^{-1}$ Fermi resonances of C-H stretching modes with combination and overtones of ring vibrations seem to cause complicated vibrational signatures and prohibit a straightforward assignment. To clarify the situation and to obtain a reliable assignment we record the spectra of ILs with different alkyl chains in this frequency range by a variety of spectroscopic methods and compare the results to anharmonic calculations of the vibrational frequencies. Beside linear Fourier transform (FT) IR and FT-Raman spectroscopy we apply also multiplex coherent anti-Stokes Raman scattering (CARS) spectroscopy using ultrashort laser pulses, which gives a different contrast due to its nonlinear character and its sensitivity to vibrational coherences and their dephasing.[34-36] In this way a better interpretation of the vibrational bands in the C-H stretching region and a molecular understanding of the interactions in these Coulomb systems is achieved.

## 2. Experimental Techniques

a) Preparation and Handling of ionic liquids (ILs)

The studied ionic liquids, which include the same anion bis(trifluoromethylsulfonyl)imide (NTf$_2^-$) but various cations i.e. 1,3-dimethyl imidazolium, 1-ethyl-3-methyl imidazolium and 1-octyl-3-methyl imidazolium, were of commercial origin (Iolitec GmbH (Denzlingen, Germany) with a stated purity of >98%. All substances were additionally dried in vacuum (p=8·10$^{-3}$ mbar) for approximately 24 hours. The water content was then determined by Karl-Fischer titration and was less than 200 ppm in all cases. Further purification was not carried out.

b) IR measurements

The Fourier transform infrared (FTIR) measurements were performed with a Bruker Vertex 70 FTIR spectrometer. The equipment for the IR measurements consists of a potassium bromide beam splitter and a room temperature DLATGS (deuterated L-alanine doped triglycene sulphate) detector with preamplifier. The accessible spectral region for this configuration lies between 1000 cm$^{-1}$ and 4500 cm$^{-1}$. An L.O.T.-Oriel variable-temperature cell equipped with calcium fluoride (CaF$_2$) windows having a path length of 0.012 mm was used for the IR measurements in transmission. For each spectrum 100 scans were recorded at a spectral resolution of 1 cm$^{-1}$.

c) Linear Raman measurements

The Bruker Vertex 70 FTIR spectrometer is equipped with an extension for Raman measurements, the RAM II FT-Raman Module. For the linear Raman measurements a

Nd:YAG laser (1064 nm) with a power of 1 W from Klastech was used. The signal was detected at a nitrogen cooled highly sensitive Germanium diode detector. This RAM II configuration provides a spectral range of 50 cm$^{-1}$ to 4000 cm$^{-1}$ and 400 scans were taken at a resolution of 1 cm$^{-1}$. For a reliable comparison, all spectra were recorded at 298 K.

d) Coherent Anti-Stokes Raman Scattering (CARS)

In addition to linear Raman measurements time resolved multiplex CARS with ultrashort excitation and narrowband probing is applied. This technique provides higher spectral contrast compared to linear Raman and allows for the characterization of dephasing kinetics of molecular vibrations.[35,36] The CARS setup is described in detail in Ref. 37. It is based on a femtosecond noncollinear optical parametric amplifier (NOPA) generating sub-50 fs Stokes pulses and a modified picosecond NOPA providing Raman pump and probe pulses with a bandwidth of about 20 cm$^{-1}$. The broadband Stokes pulses have a center wavelength of 604 nm and the pump and probe pulses are tuned to 510 nm. The pulse energies are a few hundred nJ. The three beams are arranged in a folded BOXCARS geometry for phase matching and focused into the sample to a common spot with a diameter of about 100 μm. The Stokes and the pump pulses excite vibrational coherences of Raman active modes in the CH-stretching region. The interaction of the vibrational coherences with the probe pulses results in a nonlinear polarization and the laser like CARS-signal. The signal is coupled into a spectrograph by a glass fiber and recorded by an array detector. The Raman probe can be time delayed with respect to the Stokes and pump pulses to reduce the non-resonant background and to measure the decoherence time of the vibrational excitation.

## 3. Theoretical Model

Anharmonic vibrational spectra have been calculated for a cluster model comprising four ion pairs ([$C_2$mim][$NTf_2$]) in gas phase. As shown in Fig. 1 this cluster supports a motif where all three C-H groups are hydrogen bonded to neighboring anions. Thus at least trimers are required to saturate the three potential proton donor positions at the imidazolium cation. However, in our case the bulk phase behaviour (as obtained from X-ray structures of the solid material[38]) could best described by a tetrameric ion-pair aggregate of the [$C_2$mim][$NTf_2$]. The starting point for the generation of a potential energy surface (PES) has been the equilibrium structure as obtained from a geometry optimization at the B3LYP/6-31+G(d) level of theory. The marked HBs are found to have the following bond lengths/angels: 2.03 Å/174°, C(2)-H…O, 2.47 Å/138°, C(4)-H…O, 2.75 Å/135°, C(5)-H…O. Note, that in the common terminology these HBs would be classified as being of moderate to weak strength.[39]

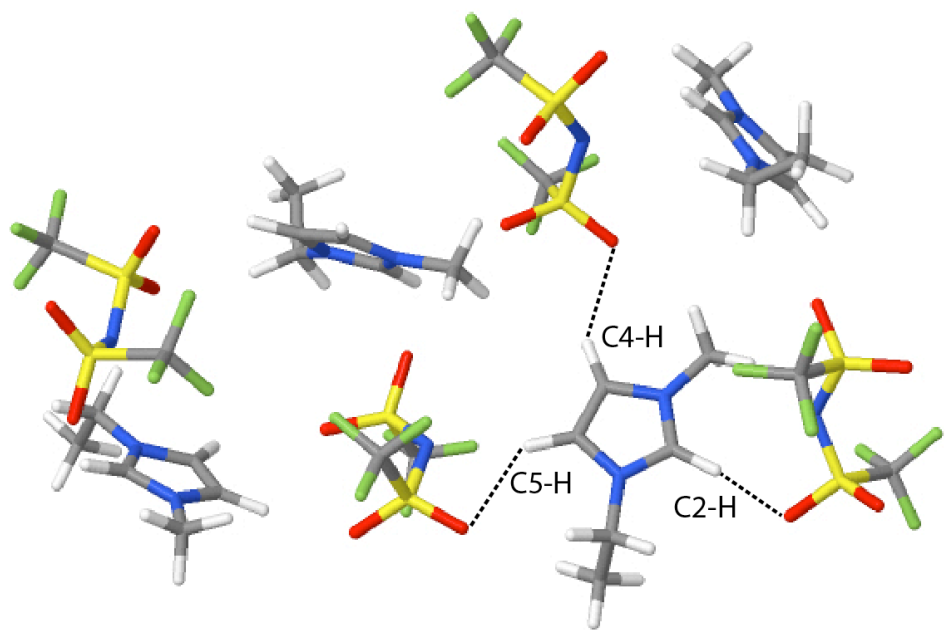

**Fig. 1:** Cluster model of [C$_2$mim][NTf$_2$] (geometry at stationary point at the B3LYP/6-31+G(d) level of theory). The three HBs are marked whose IR spectral signatures are in the focus of the anharmonic vibrational model.

For the anharmonic vibrational calculation we will use a description in terms of normal mode displacements of a set of target modes with the respective normal mode vectors spanning the potential energy surface. Since the focus is on the C-H stretching region containing the signatures of the HBs marked in Fig. 1, we have chosen the respective three normal modes shown in Fig. 2 as $Q_3$-$Q_5$. Anticipating a coupling of these modes to overtones and combination transitions of in-plane ring vibrations of the imidazolium [27, 33], we have included the two modes $Q_1$ and $Q_2$ corresponding to the modes R$_1$ and R$_2$ discussed above (cf. upper panel in Fig. 2) to arrive at a five-dimensional model, $\mathbf{Q} = (Q_1, ... Q_5)$.

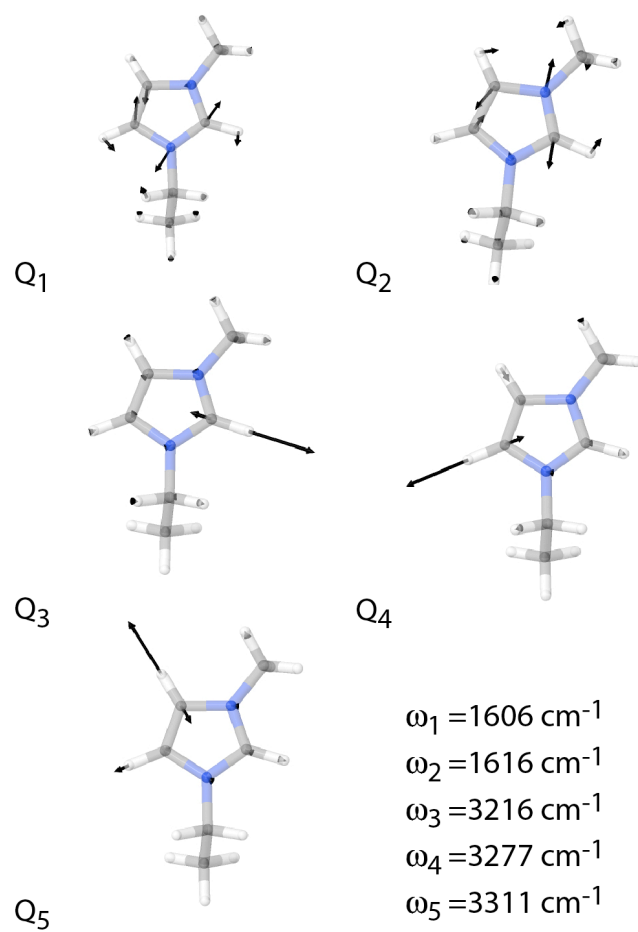

**Fig. 2:** Normal mode displacement vectors and harmonic frequencies of those coordinates that comprise the five-dimensional model used for the interpretation of the C-H stretching region of the IR spectrum of [$C_2$mim][$NTf_2$].

Neglecting rotational effects and freezing all other normal modes of the cluster at their equilibrium positions, the Hamiltonian can be written as (using dimensionless normal mode coordinates):[40]

$$H = \sum_{i=1}^{5} \frac{\hbar \omega_i}{2} \frac{\partial^2}{\partial Q_i^2} + V(\mathbf{Q}) \qquad (1)$$

In general the PES can be expressed in terms of a correlation order expansion (see, e.g., Ref. 41).

$$V(\mathbf{Q}) = \sum_i V^{(1)}(Q_i) + \sum_{i<j} V^{(2)}(Q_i, Q_j) + \ldots \qquad (2)$$

In the following we will not aim at a quantitative comparison with experiment. After all, the cluster model cannot capture all aspects of the bulk liquid. Hence we will not only restrict ourselves to an expansion up to two-mode correlations, but also include only certain two mode terms, i.e. those involving the interaction of the two ring deformation modes with the three C-H stretching vibrations. Thus we use the following approximation for the two-mode PES

$$\sum_{i<j} V^{(2)}(Q_i, Q_j) \approx \sum_{i=1}^{2} \sum_{j=3}^{5} V^{(2)}(Q_i, Q_j) \quad (3)$$

For the three vector components of the dipole moment surface (DMS), $d_{k=x,y,z}(\mathbf{Q})$, we have applied the same expansion. PES and DMS are generated on a grid comprising 14 points along the C-H stretching coordinates and 11 points for the ring deformation modes. Thus in total 844 single point calculations have been performed (B3LYP/6-31+G(d)) using Gaussian 03.[42] Subsequently, one- and two-dimensional surfaces have been fitted by a third-order spline interpolation.

Selected eigenstates and transition dipole moments are calculated as follows. First, zero-order states are determined according to the one-mode potentials

$$\left[\frac{\hbar \omega_i}{2} \frac{\partial^2}{\partial Q_i^2} + V^{(1)}(Q_i)\right] |\chi_{i,n_i}\rangle = E_{i,n_i} |\chi_{i,n_i}\rangle \quad n_i = 1,...N_i \quad i = 1,...5 \quad (4)$$

This task is accomplished by using the Fourier-Grid-Hamiltonian method using 21 grid points on the fitted PES in the ranges specified in Fig. 6 below.[43] Subsequently, this basis is used to express the two-dimensional eigenfunctions for the selected cuts of the PES, Eq. 3, i.e.

$$|\Psi_\alpha\rangle = \sum_{n_i=1}^{N_i} \sum_{n_j=1}^{N_j} C_{\alpha,n_i n_j} |\chi_{i,n_i}\rangle |\chi_{j,n_j}\rangle \quad (5)$$

Convergence for the transitions in the CH-stretching range could be obtained by choosing and $N_i = 4$ for modes $Q_1$ and $Q_2$ and $N_i = 3$ for modes $Q_3$-$Q_5$. Using the resulting $N_i \times N_j$ eigenvalues and eigenfunctions the IR transition intensities shown below are defined as

$$I(\omega) = \omega \sum_{k=x,y,z} \sum_{\alpha} |\langle\Psi_\alpha|d_k(\mathbf{Q})|\Psi_1\rangle|^2 \delta(\hbar\omega - E_1 - E_\alpha) \quad (6)$$

with the spectra drawn as stick spectra normalized to the maximum peak in the considered range. The transitions will be classified according the notation $(v_1, v_2, v_3, v_4, v_5)$, where the $v_i$ are the number of quanta in mode $Q_i$.

## 4. Experimental Results

Figure 3 shows the two spectral regions between 1500 cm$^{-1}$ and 1650 cm$^{-1}$ and between 3050 cm$^{-1}$ and 3250 cm$^{-1}$ of the FTIR spectra of the three investigated [C$_n$mim][NTf$_2$] imidazolium-based ILs with n = 1, 2 and 8. In the first region a band around 1575±15 cm$^{-1}$ is observed which exhibits a wing extending to the blue and which seems to have a double peak structure in the case of [C$_2$mim][NTf$_2$] and [C$_8$mim][NTf$_2$]. As shown below, these features can be assigned to two in-plane vibrations Q$_1$ and Q$_2$ of the imidazolium ring.

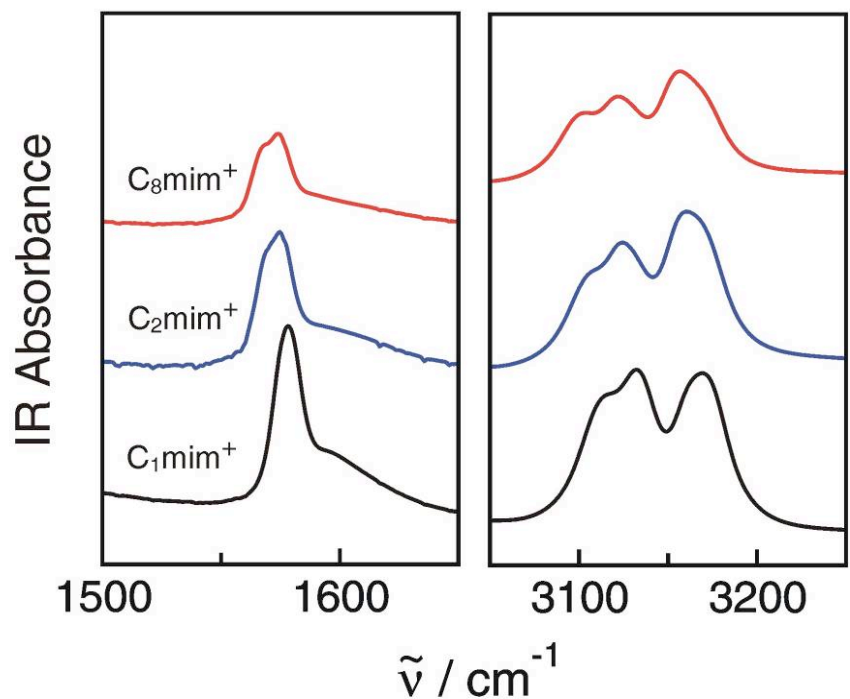

**Figure 3:** FTIR spectra of [C$_n$mim][NTf$_2$] imidazolium-based ILs with n = 1, 2 and 8. Left panel: The in-plane ring vibrational modes Q$_1$ and Q$_2$ are observed in the frequency range between 1550 cm$^{-1}$ and 1600 cm$^{-1}$. Right panel: The overtones and combinations tones of the in-plane vibrational modes along with the CH stretching vibrational modes are recorded in the frequency range between 3100 and 3200 cm$^{-1}$.

In the second region, which covers the C-H stretching contributions of the imidazolium ring the spectra of the three ILs exhibit two main features. The first one is at approximately 3120±15 cm$^{-1}$ and seems to consist in all cases of two strongly overlapping bands. The second feature appears more in the blue at approximately 3160±15 cm$^{-1}$. In the case of [C$_2$mim][NTf$_2$] and [C$_8$mim][NTf$_2$] it exhibits a shoulder in the blue wing. The assignment of these features is discussed below.

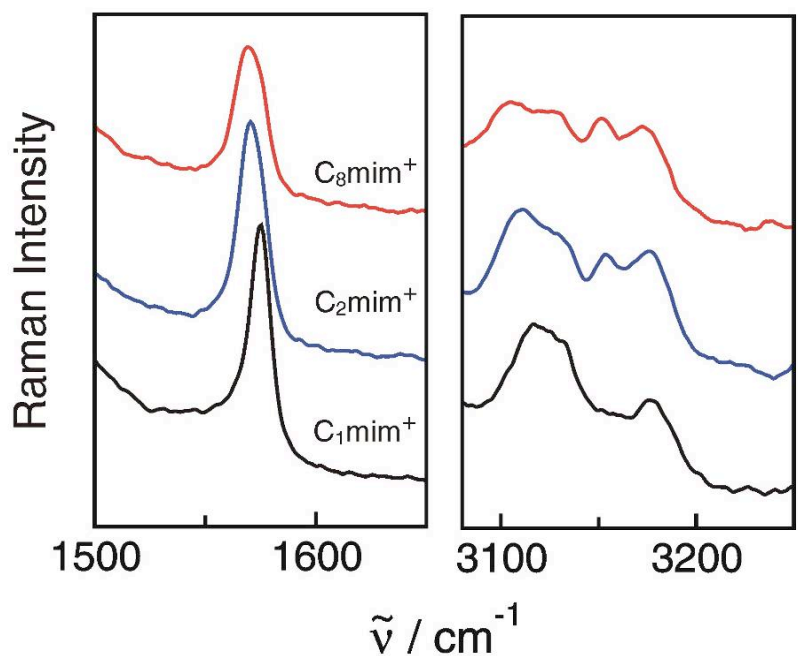

**Figure 4:** FT-Raman spectra of [$C_n$mim][NTf$_2$] imidazolium-based ILs with n = 1, 2 and 8. Left panel: The in-plane ring vibrational modes $Q_1$ and $Q_2$ are observed in the frequency range between 1550 cm$^{-1}$ and 1600 cm$^{-1}$. Right panel: The overtones and combinations tones of the in-plane vibrational modes along with the CH stretching vibrational modes are recorded in the frequency range between 3100 and 3200 cm$^{-1}$.

In Fig. 4 the Raman spectra of the three investigated ILs are presented for the two spectral regions between 1500 cm$^{-1}$ and 1650 cm$^{-1}$ and between 3080 cm$^{-1}$ and 3250 cm$^{-1}$. They resemble the IR spectra, but exhibit also some interesting differences. At 1575±15 cm$^{-1}$ a band due to the in-plane ring vibrations is observed and around 3120 cm$^{-1}$ a broad feature appears which seems to consist of two bands. In all three ILs a further band is detected at 3180±15 cm$^{-1}$ while an additional peak at 3160±15 cm$^{-1}$ can only be clearly identified in the case of [C$_2$mim][NTf$_2$] and [C$_8$mim][NTf$_2$]. Care has to be taken when comparing the intensities of the Raman bands with the IR bands in the spectral range above 3080 cm$^{-1}$ since the C-H stretching region of the alkyl groups is just below this frequency and exhibits strong Raman bands. The wings of these contributions may give rise to an underlying background and raise the spectral features of interest in a slope like manner.

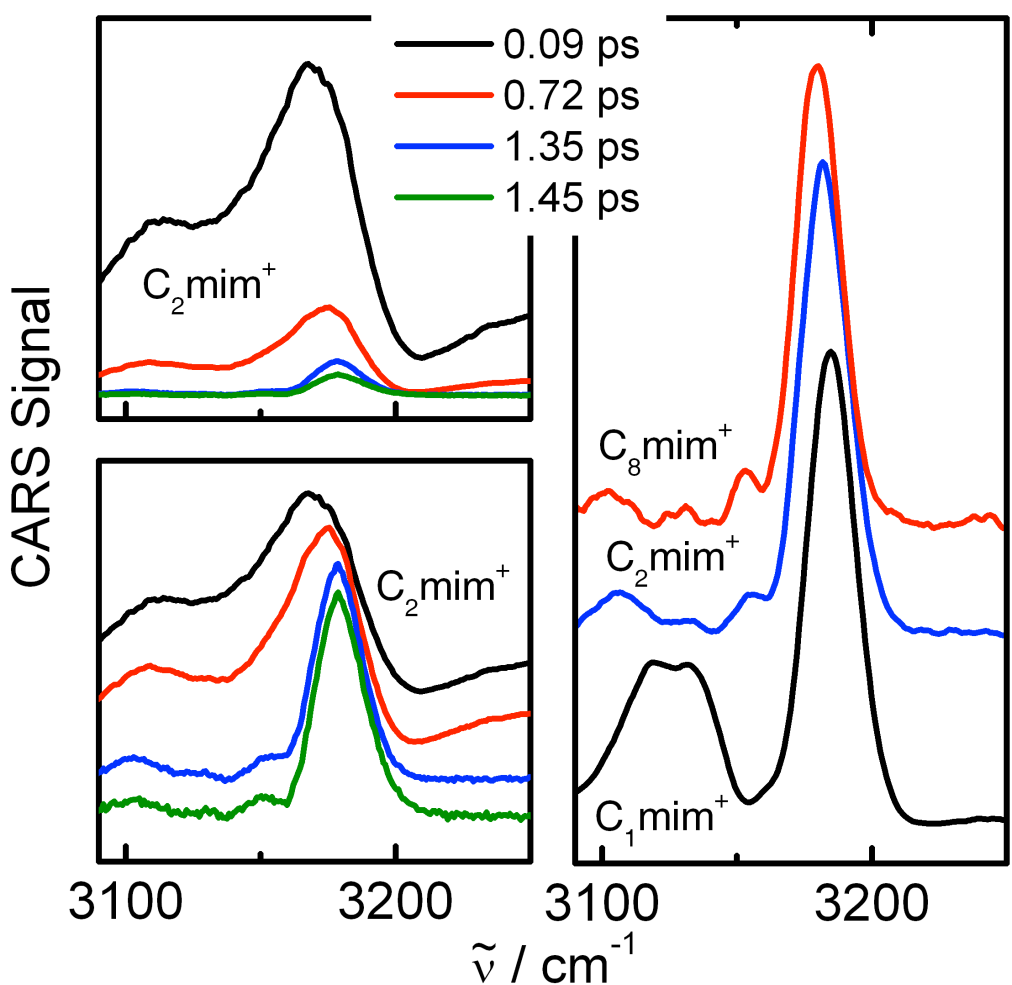

**Figure 5:** a) CARS spectra of [$C_2$mim][NTf$_2$] in the C-H stretching region of the imidazolium ring for different delay times between the CARS pump process and the probe pulse. b) Same spectra as in a) but normalized to the maximum and vertically shifted for better visibility. c) Comparison of the CARS spectra of the three considered ILs for a delay time of 1.4 ps. The spectra are normalized with respect to their maximum and vertically shifted.

Figure 5a) shows CARS spectra of [$C_2$mim][NTf$_2$] in the C-H stretching region of the imidazolium ring for different delay times between the vibrational excitation by Raman pump and Stokes pulse and the signal generation by the Raman probe pulse. The same spectra but normalized to their maxima are depicted again in Fig. 5b) for better visibility of weak features. At time zero and early delay times broad and overlapping bands are observed. With increasing time the signal decays quickly and the spectra change their shape. The structures become more pronounced, the bands adopt smaller widths and shift slightly to the blue. This behavior results from the interference between a nonresonant background and the resonant contributions.[34,36,37] The former one is due to the instantaneous electronic response of the sample and exhibits a weak frequency dependence. At a resonant contribution the difference frequency between the CARS signal and the Raman probe equals the vibrational frequency of a mode carrying Raman intensity. Since the resonant contributions are phase shifted by π/2 with respect to the nonresonant background, the vibrational signatures of the CARS spectra are strongly distorted at short delay times and the signal maxima red shifted with respect to the Raman resonances. While the nonresonant background decreases with time according to the cross correlation between Stokes and probe pulse, the resonant contributions decay after the cross correlation with half of the vibrational decoherence time.[34,35]

At delay times above 1.3 ps, where the temporal overlap between Stokes and probe pulse is negligible, the spectral positions of the observed features fit to the bands appearing in the linear Raman spectrum although the intensity ratios are quite different and the linewidths are reduced. These differences result from several effects. One reason are different dephasing and lifetimes of the distinct vibrational modes. E. g., we extracted from the time dependent CARS signal a decoherence time of 0.5 ps for the transition at 3180±15 cm$^{-1}$ whereas the other bands decay so fast that a reliable determination of the corresponding time constants is not possible. Differences between CARS and Raman spectra result also from the fact that the CARS intensity scales with the square of the Raman cross section what suppresses weak bands. In addition, the widths of bands in the CARS spectra are given by the spectral width of the probe pulse and can be smaller than the vibrational line widths.[36] This leads not only to a clearer separation of overlapping bands but also to a reduction of background signals caused by wings of strong Raman lines in the neighborhood like e. g. C-H stretching vibrations of alkyl groups which seem to increase intensity of the bands around 3120 cm$^{-1}$ in the linear Raman spectra.

A comparison of the CARS spectra of all three investigated imidazolium-based ILs is shown in Fig. 5c) for a delay time of 1.4 ps. The band at 3180±15 cm$^{-1}$ dominates the spectra and fits nicely to the corresponding feature in the Raman spectra. For [C$_2$mim][NTf$_2$] and [C$_8$mim][NTf$_2$] a weak band is observed at slightly lower wave numbers which is missing in the case of [C$_1$mim][NTf$_2$] confirming the observations of the Raman experiments. Further to the red two more weak bands appear well separated in the spectra of [C$_2$mim][NTf$_2$] and [C$_8$mim][NTf$_2$]. In the case of [C$_1$mim][NTf$_2$] they are stronger, closer together and overlap, again in agreement with the Raman spectra. Overall, the comparison between linear Raman and CARS, where weak contributions and background are suppressed, is useful in identifying the relevant features.

## 5. Results and Discussion

### 5.1 Assignment of the Recorded Vibrational Bands

Let us first discuss the feature around 3160±15 cm$^{-1}$. In the case of [C$_2$mim][NTf$_2$] and [C$_8$mim][NTf$_2$] the Raman and CARS spectra clearly indicate that it consists of two bands. In the corresponding IR spectra the band at 3160 cm$^{-1}$ exhibits a shoulder at its blue wing pointing also to two contributions but with an inversed intensity ratio compared to the CARS spectra, i.e. the band at longer wavelengths is now dominating. In the case of [C$_1$mim][NTf$_2$] all spectra reveal around 3180±15 cm$^{-1}$ only one band which is particularly clearly seen in the CARS spectrum (Fig. 5c). However, the band in the IR spectrum is red-shifted by about 10 - 20 cm$^{-1}$ with respect to the corresponding Raman and CARS band. The cation of [C$_1$mim][NTf$_2$] has two methyl groups at the imidazolium ring and exhibits a mirror symmetry whereas the two alkyl groups of the [C$_2$mim] and [C$_8$mim] cation are different from each other and the symmetry of the imidazolium ring is slightly broken (cf. Scheme 1). Therefore, it is likely that the differences between [C$_1$mim] and the other two cations in the spectral region around 3160 cm$^{-1}$ are linked to the symmetry properties of the molecules and the different selection rules of the IR and Raman transitions. In the case of [C$_1$mim][NTf$_2$] the C(4)-H and C(5)-H stretching mode are completely equivalent as long as the environment does not break the symmetry. The corresponding normal modes should therefore be symmetric and anti-symmetric combinations of the C(4)-H and C(5)-H stretching vibration. In principle both modes can be Raman as well as IR active . However, from the calculations we know that the symmetric stretch is better seen in Raman whereas the asymmetric stretch gives higher intensities in IR. Therefore we identify the mode responsible for the Raman band at 3180±15 cm$^{-1}$ with the symmetric and the mode causing the IR band at 3160±15 cm$^{-1}$ with the anti-symmetric stretching vibration. In the case of [C$_2$mim][NTf$_2$] and [C$_8$mim][NTf$_2$] the symmetry is broken and both modes should carry Raman as well as IR intensity (cf. Fig. 2). However, the disturbance of the symmetry is not very strong and the intensity distribution should still reflect the original symmetry (cf. Tab. 1). This is in accord with the observation that the Raman and CARS spectra of [C$_2$mim][NTf$_2$] and [C$_8$mim][NTf$_2$] exhibit a weak Raman band slightly red-shifted with respect to the strong band at 3180±15 cm$^{-1}$ which is missing in the [C$_1$mim][NTf$_2$] spectra while the corresponding IR spectra show a shoulder in the blue wing of the band at 3160±15 cm$^{-1}$ which is absent in the case of [C$_1$mim][NTf$_2$]. Both bands are red-shifted with respect to the frequency of a free C(4/5)-H stretching mode of the imidazolium ring.[27] This indicates that both C-H groups are involved in moderate HBs as indicated in Fig. 1.

The situation is different for the feature around 3120±15 cm$^{-1}$ which consists of two contributions in the case of all investigated ILs and independent of the applied spectroscopic method. So it cannot simply be assigned to the C(2)-H stretching mode. The IR and Raman spectra reveal strong features around 1575±15 cm$^{-1}$ due to in-plane ring modes indicating that nearby 3120 cm$^{-1}$ overtones of these modes are expected. However, they should carry only little intensity and it would be surprising if the corresponding bands are of comparable strength as the fundamentals of the C-H stretching vibrations. But the overtones can mix with the C-H stretching modes and form Fermi resonances if the difference in frequency is small and the corresponding PES anharmonic (see below). Therefore we conclude that the double structure results from the C(2)-H stretching vibration and possible Fermi resonances of C-H modes with overtones of the in-plane ring modes. Accordingly, the C(2)-H stretching mode is more red-shifted than the C(4)-H and C(5)-H stretching vibrations. This provides strong evidence that the C(2)-H group is involved in a stronger but still moderate HB.

## 5.2 Comparison to Simulated Spectra

In the following we show that the assignment given above is also strongly supported by simulations of the spectra, which take anharmonicities into account. There are two effects shaping the IR spectrum in the CH-stretching range. First, the diagonal anharmonicity, which is partly due to H-bonding. Second, the Fermi resonance interaction of the CH-stretching fundamental and ring deformation overtone transitions. In Fig. 6 we show the anharmonic one-mode potentials for two representative modes together with the probability densities of the zero-order eigenstates. For the ring deformation mode $Q_2$ we notice a pronounced harmonic character (same holds for $Q_1$, not shown). The fundamental transition is calculated at 1621 cm$^{-1}$, which almost agrees with the harmonic value of 1616 cm$^{-1}$. For $Q_1$ we have 1611 cm$^{-1}$ versus the harmonic value of 1606 cm$^{-1}$. Comparing these two values with the experimental spectrum, we observe a blue-shift. Since the experimental assignment is considered to be certain (cf. Fig. 3), we have scaled the two normal mode coordinates such as to obtain an agreement with respect to the fundamental ring deformation transitions. Since $Q_1$ and $Q_2$ have an anharmonic IR intensity ratio of 0.75 a scaling of $Q_2$ to the lower frequency peak at 1569 cm$^{-1}$ and of $Q_1$ to the higher frequency peak at 1574 cm$^{-1}$ has been performed. In passing we note that this scaling is in the range of what is usally applied when using DFT-based harmonic frequencies.

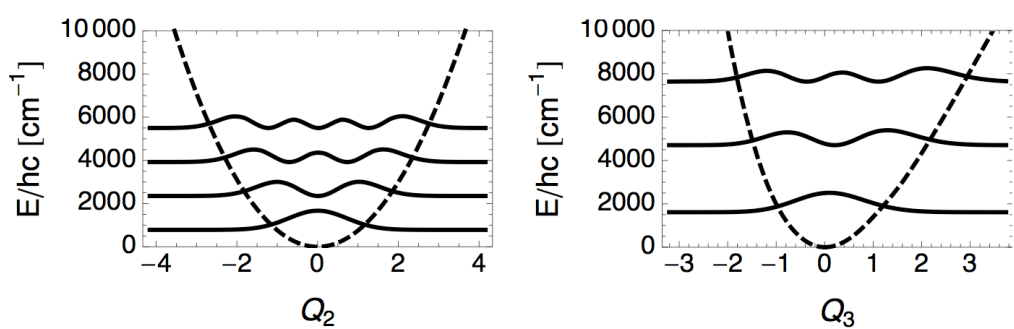

**Fig. 6:** One-dimensional potentials (dashed) and probability densities for the lowest eigenstates (solid, Eq. 4) of modes $Q_2$ and $Q_3$ (dimensionless coordinates, grid boundaries for other modes are chosen accordingly).

For the C(2)-H-stretching mode, $Q_3$, a pronounced anharmonicity is observed in Fig. 6 causing a red-shift from the harmonic transition frequency of 3216 cm$^{-1}$ down to 3093 cm$^{-1}$. Similar albeit smaller red-shifts are found for the other C-H stretching fundamental transitions.

| $(v_1,v_2,v_3,v_4,v_5)$ | harmonic | $V^{(1)}(Q_i)$ | $V^{(2)}(Q_1,Q_3)$ | $V^{(2)}(Q_1,Q_4)$ | $V^{(2)}(Q_2,Q_3)$ | experiment |
|---|---|---|---|---|---|---|
| (0,0,0,0,1) | 3311 (0.11) | 3224 (0.03) | - | - | - | 3173 |
| (0,0,0,1,0) | 3277 (0.37) | 3189 (0.31) | - | 3201 (1.00) | - | 3158 |
| (2,0,0,0,0) | 3148 (0.00) | 3150 (0.00) | 3148 (0.05) | 3138 (0.06) | - | 3125 |
| (0,2,0,0,0) | 3138 (0.00) | 3140 (0.00) | - | - | 3141 (0.12) | 3125 |
| (0,0,1,0,0) | 3216 (1.00) | 3093 (1.00) | 3090 (1.00) | - | 3087 (1.00) | 3104 |

**Table 1**: Harmonic and anharmonic frequencies (in cm$^{-1}$) as well as normalized intensities (in parentheses), calculated using diagonal anharmonicities only ($V^{(1)}(Q_i)$) as well as various cuts through 2D potentials ($V^{(2)}(Q_i,Q_j)$). Note that harmonic and 1D anharmonic transitions for the ring deformation modes are given after scaling of the fundamental transition (see text). Only intensities exceeding 0.01 are reported. The experimental values are taken from the IR spectrum.

The results of the anharmonic calculations are compiled in Tab. 1. As far as the overtone transitions of the ring deformation modes at 3150 cm$^{-1}$ and 3140 cm$^{-1}$ for $Q_1$ and $Q_2$, respectively, are concerned we notice that the effect of diagonal anharmonicity (mechanical and electronical) is essentially negligible, i.e. the overtone transitions have no noticeable intensity.

Next we discuss the effect of a Fermi resonance interaction based on selected PES and DMS cuts. Here we set the focus on those mode combinations, which give rise to some intensity in the ring deformation overtone transitions, see Tab. 1. This includes the interactions between the pairs ($Q_1,Q_3$), ($Q_1,Q_4$), ($Q_2,Q_3$), what can be rationalized by looking at the atomic displacements in Fig. 2. The main conclusion to be drawn from Tab. 1 is that Fermi resonance interaction between *both* the C(2)-H and the C(5)-H stretching fundamental and the overtone transitions of the two ring deformation modes influences the absorption spectrum in the considered range.

The present theoretical results combine previous assessments of this spectral range which had a focus either on H-bonding in a cluster model[33] or on Fermi-resonance interaction in isolated imidazolium.[27] The effect of both, H-bonding and condensed phase "packing", can be estimated by comparison with the gas phase imidazolium cation studied in Ref. 27. These authors report anharmonic frequencies of 3181 cm$^{-1}$ for the C(2)-H fundamental and 3196 and 3176 cm$^{-1}$ for the in-phase and out-of-phase C(4,5)-H vibrations, respectively. Compared to the results given in Tab. 1 we notice that the situation is not clear-cut for the C(4,5)-H vibrations, but for the C(2)-H case a clear difference is observed and the red-shift of about 90 cm$^{-1}$ can be predominantly assigned to be an effect of H-bonding to the O atom of the nearest NTf$_2$ anion. In passing we note that this is in accord with the stronger acidity of the C(2)-H site as compared with C(4/5)-H. Hence, the present calculations of anharmonic vibrational spectra performed for a [C$_2$mim][NTf$_2$] cluster provide evidence for the combined effect of H-bonding and Fermi-resonance interactions. Only after inclusion of the diagonal anharmonicity the CH-stretching fundamentals are found in the range observed in the experiment. Further, accounting for the Fermi resonance coupling gives rise to a peak in the gap between the C(2)-H and the C(4/5)-H stretching fundamental transitions. Although the results are in semi-quantitative agreement with experiment, the present simple model does not provide the correct intensities. Inspecting the results obtained after inclusion of the diagonal anharmonicity in Tab. 1 we notice that the overtone transitions of the ring deformation modes are essentially in the middle of the gap formed by the C(2)-H and C(5)-H fundamentals. Having a gap of about 40 cm$^{-1}$ to 50 cm$^{-1}$, the conditions for Fermi resonance are rather unfavorable. Here slight changes might have a substantial effect on the oscillator strength. In principle one could attempt to scale the CH-stretching fundamentals with the goal of reproducing the spectra. While this procedure is applicable to the well assignable ring deformation fundamental transitions, it would render the model to become empirical in the CH-stretching region.

## 7. Summary and Conclusions

The three different imidazolium-based ionic liquids [C$_n$mim][NTf$_2$] with n = 1, 2 and 8 have been studied applying IR, linear Raman, and multiplex CARS spectroscopy and the results are compared to simulated anharmonic vibrational spectra based on five modes of a cluster consisting of four [C$_2$mim][NTf$_2$] ion pairs. The Raman band at 3180±15 cm$^{-1}$ and the IR band at 3160±15 cm$^{-1}$ are assigned to the more or less symmetric and anti-symmetric combination of the C(4)-H and C(5)-H stretching vibration of the imidazolium ring. The feature around 3120±15 cm$^{-1}$ consists of two bands and results from the C(2)-H stretching mode and Fermi resonances of the C-H stretching vibrations with overtones of in-plane ring deformations. The calculations indicate that this feature cannot stem from pure overtones, since the potential energy surfaces of the ring deformation modes are both nearly harmonic and their overtones should not have any intensity by themselves. In addition, the Fermi resonances are mainly due to coupling with the C(2)-H stretching mode. Our results strongly support the important role played by hydrogen bonding in ILs. In particular, it was found that the C(4)-H and C(5)-H groups are involved in weak HBs while the hydrogen bond of the C(2)-H group is somewhat stronger.

In conclusion, it was demonstrated that the combination of various vibrational spectroscopic techniques and anharmonic frequency calculations allows to disentangle congested vibrational signatures.

## Acknowledgement

This work was supported by the Deutsche Forschungsgemeinschaft (DFG) within the SFB 652.